\documentclass[11pt]{article}

\usepackage{aas_macros,amsmath,amssymb,comment,cite,esint,graphicx,mathtools}
\usepackage[margin=.8in,letterpaper]{geometry}
\usepackage[colorlinks=true]{hyperref}
\usepackage[affil-it]{authblk}
\usepackage{subcaption}
\usepackage[utf8]{inputenc}
\usepackage{mathrsfs}
\usepackage{appendix}
\usepackage{amssymb}
\usepackage{float}
\usepackage{color}
\usepackage{cite}
\usepackage{hyperref}
\hypersetup{pageanchor=false}
\usepackage{indentfirst}
\usepackage{url}
\usepackage{float}
\usepackage{caption}
\usepackage[numbers,square,comma,sort&compress,merge]{natbib}
\usepackage{esint}
\usepackage{overpic}
\usepackage{graphicx}
\usepackage{epsf,amsmath,bbold,amsfonts,stmaryrd}
\usepackage{textcomp}
\usepackage{ulem}
\usepackage{tikz}

\numberwithin{equation}{section}
\let\originalleft\left
\let\originalright\right
\renewcommand{\left}{\mathopen{}\mathclose\bgroup\originalleft}
\renewcommand{\right}{\aftergroup\egroup\originalright}
\mathcode`\*="8000
{\catcode`\*=\active\gdef*{\mathclose{}\,\mathopen{}}}

\newcommand{\td}[1]{\tilde{#1}}

\newcommand{\be}{\begin{equation}}
\newcommand{\ee}{\end{equation}}
\newcommand{\bea}{\setlength\arraycolsep{2pt} \begin{eqnarray}}
\newcommand{\eea}{\end{eqnarray}}
\newcommand{\nn}{\nonumber}

\newcommand{\mA}{{\mathcal A}}
\newcommand{\mE}{{\mathcal E}}

\newcommand{\mB}{\mathfrak B}
\newcommand{\mS}{{\mathcal S}}
\newcommand{\mT}{{\mathcal T}}
\newcommand{\mO}{{\mathcal O}}
\newcommand{\ma}{\mathsf{a}}
\newcommand{\mb}{\mathsf{b}}
\newcommand{\mc}{\mathsf{c}}
\newcommand{\md}{\mathsf{d}}

\def\d{\delta}
\def\D{\Delta}
\def\f{\frac}

\def\G{\Gamma}
\def\lm{\lambda}
\def\l{\left}
\def\L{\Lambda}
\def\m{\mu} 
\def\n{\nu} 
\def\nn{\nonumber}
\def\pl{\partial}
\def\p{\phi} 
\def\td{\tilde} 
\def\r{\right}

\def\t{\theta}
\def\T{\Theta}

\def\ep{\epsilon}

\def\be{\begin{equation}}
\def\ee{\end{equation}}
\def\bag{\begin{aligned}}
\def\eag{\end{aligned}}
\def\bea{\begin{eqnarray}}
\def\eea{\end{eqnarray}}
\def\ba{\begin{array}}
\def\ea{\end{array}}

\def\bc{\begin{center}}
\def\ec{\end{center}}

\begin{document}
\title{Electromagnetic effects on charged particles in NHEK}

\author{
Yehui Hou$^{1}$, Zhenyu Zhang$^{1}$, Minyong Guo$^{2\ast}$,
Bin Chen$^{1, 3, 4}$}
\date{}

\maketitle

\vspace{-10mm}

\begin{center}
{\it
$^1$Department of Physics, Peking University, No.5 Yiheyuan Rd, Beijing
100871, P.R. China\\\vspace{4mm}

$^2$ Department of Physics, Beijing Normal University,
Beijing 100875, P. R. China\\\vspace{4mm}

$^3$Center for High Energy Physics, Peking University,
No.5 Yiheyuan Rd, Beijing 100871, P. R. China\\\vspace{4mm}

$^4$ Collaborative Innovation Center of Quantum Matter,
No.5 Yiheyuan Rd, Beijing 100871, P. R. China\\\vspace{2mm}
}
\end{center}

\vspace{8mm}

\begin{abstract}
We investigate the motions of charged particles in the near horizon region of an extreme Kerr black hole with weak electromagnetic fields. There is an enhanced symmetry in the NHEK  geometry. We find that when the electromagnetic field respects this enhanced symmetry, which we refer to as the maximally symmetric electromagnetic (MSEM) field, the equations of motion of charged particles get simplified into a set of decoupled first-order differential equations.  We discuss the motions of charged particles in two MSEM fields, one being the force-free field and the other being the vacuum fields. Even though the radial motions are similar to the geodesics in NHEK geometry, the angular motions could be affected by the electromagnetic field significantly. In particular, for the vacuum solution which is  produced by a weakly charged black hole, there exist stable vortical motions if the electromagnetic parameter is above the critical value  $\mB_c = \sqrt{3}$.  These vortical motions do not cross the equatorial planes, and the charged particles in them radiate non-thermally. We discuss the corresponding astrophysical implications.

\end{abstract}

\vfill{\footnotesize $\ast$ Corresponding author: minyongguo@bnu.edu.cn}

\maketitle

\newpage
\baselineskip 18pt
\section{Introduction}\label{sec1}

Black holes are astronomical objects of great interests. In recent years, the horizon-scale images of supermassive black holes released by Event Horizon Telescope (EHT) collaborations \cite{EventHorizonTelescope:2019dse, EventHorizonTelescope:2022wkp} have pushed the study of black hole physics in strong gravity regime  to a new level.  By comparing the observational data with theoretical models, they allow us to estimate the masses and spins of  black holes, draw the magnetic field configurations around the black holes and investigate the physics in the accretion disks. 

The (electro-)magnetic field  around a black hole affect significantly the images of the black hole.  It not only changes the worldlines of charged particles from the geodesics to more general trajectories, but also induces the synchrotron radiation. The Lorentz force significantly affects the motions of electrons and ions, which triggers a series of observable phenomena such as the jets, the synchrotron radiation, and the gamma-ray bursts \cite{Rueda:2022fgz}. The polarized  images of black holes encode rich information on the structure of the electromagnetic fields and the dynamics of charged particles\cite{EventHorizonTelescope:2021srq, Hu:2022sej, Zhu:2022amy, Lee:2022rtg, EventHorizonTelescope:2021btj, Gelles:2021kti, Qin:2022kaf, Liu:2022ruc}. The electromagnetic fields surrounding the astronomical black holes \cite{Dovciak:2004gc, Piotrovich:2010aq, Daly:2019srb} are widely believed to originate from  the accreted plasma, even though there are also discussions about the possibility that the black hole charge act as the source of electromagnetic fields\cite{Zajacek:2018ycb, Zajacek:2019kla}. In any case,  there are new configurations of classical electrodynamics fields in the curved spacetime,  which do not exist in a flat spacetime, and the motions of the charged particles in them present novel features. 

For geometrically thin accretion disks, the fluid motion can be approximated at the lowest order as single-particle trajectories on the equatorial plane. The problem of the fluid motion  effectively reduces to an analysis of the effective radial potential of a single charged particle \cite{Prasanna:1978vh, Aliev:2002nw, Frolov:2010mi, Abdujabbarov:2009az, Hussain:2014cba, Shaymatov:2021qvt, Zahrani:2022fdd}. Moreover, richer physical processes exist in the off-equatorial region governed by the magnetosphere of black holes. In this region, the single-particle approximation of the dilute plasma works even better \cite{Zajacek:2018ycb}. For integrable systems like the Reissner-Nordström and Kerr-Newman black holes, the charged particle motions are regular \cite{Carter:1968rr, Pugliese:2011py, Hackmann:2013pva}, and behave as continuous lines on the Poincare section. However, the electromagnetic fields generically destroy the integrability, making the analytical study of the off-equatorial motions impossible. One has to employ the numerical methods and carefully select the initial conditions. More related works can be found in \cite{Kovar:2008ta,Kovar:2010ty,Shiose:2014bqa,Tursunov:2016dss}. In addition, the non-integrability generally leads to chaotic motions of the particles, as the scattered dots on the Poincare section \cite{Takahashi:2008zh, Kopacek:2010at, Stuchlik:2015nlt, Liu:2018bmn, Sun:2021ndd}. However, if the spacetime is highly symmetric, and for special electromagnetic field configuration, the off-equatorial motions of charged particles can still be studied analytically, which allow us to gain  more insights of the physics.

It is well known that  the near-horizon-extreme-Kerr (NHEK) geometry has an enhanced symmetry,  which has led to extensive theoretical studies \cite{Bardeen:1999px, Guica:2008mu, Bredberg:2009pv, Chen:2010fr, Kapec:2019hro, Compere:2020eat, Galajinsky:2010zy, Galajinsky:2011xp}. Based on observational evidences of high-spin black holes \cite{McClintock:2006xd,Gou:2013dna}, there is a variety of analytical studies on the possible astronomical observation \cite{Grumiller:2009gf,AlZahrani:2010qb, Compere:2015pja, Gralla:2016jfc, Camilloni:2020qah, Camilloni:2020hns, Porfyriadis:2016gwb,Compere:2017zkn, Gralla:2017ufe, Guo:2018kis, Guo:2019lur, Li:2020val,Yan:2021ygy, Yan:2019etp}. Among them, the force-free electrodynamics in NHEK were studied in \cite{Compere:2015pja, Gralla:2016jfc, Camilloni:2020qah, Camilloni:2020hns}, and the extreme Kerr accretion was investigated in \cite{Grumiller:2009gf, Compere:2017zkn}. In \cite{AlZahrani:2010qb}, the authors studied the motions of charged particles under the Wald potential in the NHEK geometry. They found that the motions can be solved analytically, and they investigated the motions in the equatorial plane and along the spin axis. However, the  electromagnetic effects on the off-equatorial motions of the charged particles in the extreme Kerr spacetime has not been explored much. 

In this work, we study the motions of charged particles in the NHEK geometry, in the presence of weak electromagnetic fields. We neglect the backreaction effect to the spacetime, as the electromagnetic field is weak. We find that under maximally symmetric electromagnetic (MSEM)  fields, the Carter constant is geometrically-realized by the symmetry of the system. Thus, we are allowed to have an analytical study of the motion of charged particles. We derive the first-order differential equations of motion of charged particles for universal MSEM fields and discuss some general properties of the angular potential. Then we consider two kinds of MSEM fields,  the force-free one and a general vacuum field, and study the motion of charged particles in them. In particular, the vacuum field with the $Z_2$ symmetry under $\t \rightarrow \pi-\t$ is exactly the Wald solution \cite{Wald:1974np} in the near-horizon limit, which describes an electromagnetic field generated by a weakly charged black hole. We present a complete analysis of the charged particle motions in the Wald solution, and especially have a detailed discussion on the vortical motions.

The paper is organized as follows. In Sec. \ref{sec2}, we shall first review the NHEK geometry and its isometric group. In Sec. \ref{sec3}, we derive the first-order differential equations of motion of charged particle motion under MSEM fields. The fields do not affect the radial motion, and we briefly discuss its classification. In Sec. \ref{sec4}, the angular motions under different types of MSEM fields are explored. We summarize and conclude this work in Sec. \ref{sec5}. We work in the geometrized unit with $G = c = 1$ in this paper.

\section{Extreme Kerr black holes}\label{sec2}

\subsection{NHEK limit}

The Kerr geometry is a generally accepted description of astrophysical black holes within the framework of Einstein's gravity. In Boyer-Lindquist (BL) coordinates, the metric is given by
\be
ds^2=-\f{\D}{\Sigma}\big(d\td{t}-a\sin^2{\td{\t}}d\td{\phi}\big)^2+\Sigma\l(\f{d\td{r}^2}{\D}+d\td{\phi}^2\r)+\f{\sin^2{\td{\t}}}{\Sigma}\big[ad\td{t}-(\td{r}^2+a^2)d\td{\phi}\big]^2\, ,
\label{BL}
\ee
where 
\bea
\Sigma(\td{r},\td{\t})=\td{r}^2+a^2\cos^2\td{\t}\,,\quad\Delta(\td{r})=\td{r}^2-2M\td{r}+a^2\,.
\eea 
The parameter $M$ is the mass of the black hole, $a = J/M$ denotes the spin parameter, with $J$ the angular momentum of the black hole. The metric is invariant under $a \rightarrow -a$, $\p \rightarrow -\p$, so we can set $a\geq 0$ for simplicity and without loss of generality. The coordinate singularities are located at $\td{r}_{\pm}=M\pm\sqrt{M^2-a^2}$, and $\td{r}_+$ is the radius of the event horizon. To avoid naked singularities, $a \leq M$ has to be satisfied. When $a \rightarrow M$, one has an extreme Kerr black hole, whose near-horizon geometry is very different from the usual Kerr spacetime. In this case, one can use the method developed by Bardeen and Horowitz \cite{Bardeen:1999px} to obtain the near-horizon geometry. Consider a small deviation from the extremality
\bea
a = M\sqrt{1-\ep^2}\,,
\eea
where $\ep \ll 1$ is a small factor. Then we perform a dilation onto the horizon
\bea
\label{trans}
t =\f{\ep^p }{2M} \td t \, , \ \ r = \f{\td r -M}{\ep^pM} \, ,\ \ \t = \td \t \, ,\ \ \phi = \td \phi - \f{\td t}{2M}\,,
\eea
with $0<p\leq 1$ being the ``zoom power''. Different $p$ will zoom into different near-horizon regions (e.g., the ISCO is in $p = 2/3$), and the proper distance is proportional to $|\log{\ep}|$ \cite{Kapec:2019hro}.
For each $p < 1$, expanding the metric to the leading order of $\ep$, we can get the near-horizon-extreme-Kerr (NHEK) geometry
\bea\label{NHEK1}
ds^2 = 2M^2\G \bigg[ -r^2dt^2+\f{dr^2}{r^2}+d\t^2+\L^2(d\phi+rdt)^2 \bigg] + \mathcal{O}(\epsilon) \, ,
\eea
\bea
\G=\f{1+\cos^2{\t}}{2} \, , \quad \L = \f{2\sin{\t}}{1+\cos^2{\t}} \,.
\eea
It is worth emphasizing that the NHEK geometry is an exact solution to the vacuum Einstein equations and describes the near-horizon region of the extreme Kerr black hole through the coordinate rescaling Eq. (\ref{trans}). However, it is not asymptotically flat. For a fixed $\theta$, the NHEK geometry is a $U(1)$-fibred AdS$_2$, in which  the geometry along the spin axis is exactly $AdS_2$. For $p=1$, the expansion gives the near-NHEK geometry at the leading order, which takes the form
\bea\label{near}
ds^2 = 2M^2\G \bigg[ -(r^2-1)dt^2+\f{dr^2}{r^2-1}+d\t^2+\L^2(d\phi+rdt)^2 \bigg] + \mathcal{O}(\epsilon) \, .
\eea
Since $p$ corresponds to the extent of zooming, near-NHEK is the deepest region where the horizon is located. The other $p<1$ corresponds to the regions with different depths, which are all described by the NHEK. As pointed out in \cite{Bredberg:2009pv}, Eq.~\eqref{near} and Eq.~\eqref{NHEK1} are locally diffeomorphic to each other, under a coordinate transformation not involving $\t$-direction. On the other hand, as shown in Sec. \ref{sec31}, the electromagnetic field does not change the radial motion of the particles. Thus we only discuss physical processes in the NHEK geometry without loss of generality. From now on, we set $M = 1$ for simplicity.

\subsection{The isometry group}

Different from stationary and axisymmetric Kerr spacetime \eqref{BL}, which has two Killing vectors $\pl_{\td{t}}$, $\pl_{\td{\p}}$, the NHEK geometry possesses an enhanced Killing symmetry. At an infinitesimal level, the isometry group is generated by
\bea\label{Killing}
&& W = \pl_{\phi} \, , \quad
H_0 = t\pl_t-r\pl_r \, , \nn \\
&& H_+=\pl_t \, , \quad
H_- =\bigg( t^2+\f{1}{r^2}\bigg)\pl_t - 2tr\pl_r-\f{2}{r}\pl_{\phi} \, .
\eea
The Killing vectors $W$ and $H_+$ are the translations along $\p$-direction and $t$-direction, respectively, and $H_0$ implies a self-similarity under the scaling $t\rightarrow t/\lambda \, , \, r \rightarrow \lambda r$. The fourth Killing vector $H_-$ denotes the time translational invariance when transforming to the global coordinates (an analog) of $AdS_2$ \cite{Bardeen:1999px}. The commutation relations between the generators are
\bea
[H_0, H_{\pm}] = \mp H_{\pm} \, , \ \ [H_+, H_-] = 2H_0 \, , \ \ [W, H_{\pm,0}]=0 \, ,
\label{commutation}
\eea
which generate a $\mathsf{SL}(2,\mathbb{R})\times \mathsf{U}(1)$ isometry group. Simply speaking, the $\mathsf{SL}(2,\mathbb{R})$ subgroup is the isometry of AdS$_2$, while $\mathsf{U}(1)$ is the translational group along $\phi$. The quadratic Casimir is
\be\label{casimir}
C^{\m\n} = -H^{\m}_0H^{\n}_0 + \f{1}{2}(H^{\m}_+H^{\nu}_-+H^{\m}_-H^{\nu}_+) \, .
\ee
Recall that the Kerr spacetime has an irreducible Killing tensor, $K^{\mu\nu}$, which generally cannot be constructed using any Killing vectors. Thus, the Killing tensor implies a non-geometrically-realized symmetry and a Carter constant of motion along geodesics, $K = K^{\mu\nu}p_{\mu}p_{\nu}$. As a result, $r$ and $\t$ in the Hamilton-Jacobi equation can be separated\cite{Carter:1968rr}. For an extreme Kerr black hole, when zooming into NEHK, $K^{\mu\nu}$ becomes reducible
\be
K^{\m\n} = g^{\m\n} + W^{\m}W^{\n} + C^{\m\n} \, ,
\ee
where $g^{\m\n}$ is the metric component. Therefore, due to the enhanced Killing symmetry in NHEK, $K^{\m\n}$ can be constructed by the Killing vectors, and the Carter constant is geometrically-realized by the Casimir. This property will help us to obtain the equations of motion in NHEK with electromagnetic fields.

\section{Particle dynamics in MSEM field}\label{sec3}

In this section, we study the dynamics of charged particles in NHEK with a weak electromagnetic field, whose backreaction to the spacetime can be neglected. Despite the Lorentz force, the Killing symmetries can still be used to construct conserved quantities along the trajectories in the presence of a maximally symmetric electromagnetic (MSEM)  field, whose Lie derivatives along the Killing vector fields of NHEK are vanishing. Consequently, we may obtain the first-order differential equations of motion of charged particles.

\subsection{The equations of motion}\label{sec31}
If the specific charge is large enough, the Lorentz force can be rather strong compared to the gravitation. For an electrons, even a magnetic field of 1 Gauss around a black hole of solar mass can produce an Lorentz force of nearly $100$ times magnitude of the gravitation. For a particle of charge $q$ and mass $m$, neglecting the radiation-reaction force, its equation of motion reads
\bea
D_{\tau} U^{a} =\f{q}{m} F^{ab} U_{b} \, ,
\eea
where $U^{a} $ and $F^{ab}$ are the 4-velocity and electromagnetic field strength tensor, respectively, and $D_{\tau}$ is the covariant derivative about the proper time $\tau$. If the electromagnetic field is invariant along a Killing vector $\xi$, i.e., 
\bea
\mathcal{L}_{\xi}\mA= [\xi, \mA]=0\,,
\eea
where $\mA$ is the electromagnetic field potential, $\mathcal{L}_{\xi}$ is the Lie derivate along $\xi$, then we have
\bea
D_{\tau}(\xi_{\m}P^{\m}) = q U^{\n}( \mathcal{L}_{\xi} \mA)_{\n} = 0 \, .
\eea
Here $P^{\m} = mU^{\m} + q A^{\m}$ denotes the canonical momentum. Even though one can define the  energy and angular momentum of a charged particle in a stationary and axisymmetric spacetime, one has to evolve the second-order differential equations of motion numerically\cite{Takahashi:2008zh, Liu:2018bmn} since the Carter constant gets lost. However, in NHEK,  the geometric symmetry of the electromagnetic field can be enhanced in such a way that  the dynamics of the charged particles could be simplified. More precisely, if the electromagnectic field is maximally symmetric, it has a $\mathsf{SL}(2,\mathbb{R})\times \mathsf{U}(1)$ invariant potential satisfying
\bea
\mathcal{L}_{W}\mA=0 \, , \, \quad \mathcal{L}_{H_0}\mA=0 \, , \, \quad \mathcal{L}_{H_{\pm}}\mA=0 \, , \label{MS}
\eea
and there are four constants of motion,
\bea
&& E = -H_+^{\m}P_{\m} = -P_t \, , \quad L = W^{\m}P_{\m} = P_{\phi} \, ,\nn \\
&& h_0 = H_0^{\m}P_{\m} = -Et - rP_r \, , \quad h_- = H_-^{\m}P_{\m} = -\bigg( t^2+\f{1}{r^2}\bigg)E - 2trP_r-\f{2}{r}L \, .
\eea
The constants $E$ and $L$ are the NHEK analogues of the energy and angular momentum\footnote{By the rescaling Eq.~\eqref{trans} we have $\widetilde{E} = \epsilon E/2 + L/2$, $\widetilde{L} = L$, where the tilde denotes the quantities in the Kerr spacetime. Thus, when moving in the NHEK region, the particle's Kerr energy is determined by $L$ if neglecting $\mO(\epsilon)$ correction. Note that only particles with $L>2m$ can reach the infinity.}. The constant $h_0$ originates from the self-similarity of the system under $t\rightarrow t/\lambda \, , \, r \rightarrow \lambda r$, and $h_-$ is from the translational invariance along the global time. An important property is that under an MSEM field  the Carter constant is restored
\bea
C = C^{\m\n}P_{\m}P_{\n} =-h_0^2 - E h_- \, ,
\eea
which implies the separability of the Hamilton-Jacobi equation. To see this, we need the specific form of the MSEM field. Denote the potential as $\mA_{\text{MS}}$, in the gauge $A_{\t} = 0$, the 1-form potential satisfying Eq.~\eqref{MS} must be of the form
\be\label{max sym}
\mA_{\text{MS}} =f(\t)(d\phi+rdt),
\ee
where $f(\theta)$ is an undetermined function of $\t$. We may transform the 1-form potential into a vector potential with only one nonvanishing component
\be
A^\phi=\f{f(\t)}{2\G\L^2}.
\ee
In other words,  the vector potential $\mA_{\text{MS}} $ is proportional to $W$, such that it is obviously invariant under the isometry group. Within the gauge we have chosen, $C$ directly gives a first-order differential equation,
\be\label{casimir2}
C = -r^2 U_r^2+\f{E^2}{r^2} + \f{2EL}{r} \, ,
\ee
where and hereafter $E,L$ are rescaled by $1/m$, and $C$ rescaled by $1/m^2$, thus all the conserved quantities become dimensionless. Combining Eq. (\ref{casimir2}) with $U^a U_a=-1$, we can get rid of the $U_r$ component and find a first-order differential equation of the polar angle,
\be
C = U^2_{\t} + \bigg( \f{1}{\L^2}-1 \bigg)L^2 + 2\Gamma +\f{k^2f^2(\t)-2k Lf(\t)}{\L^2} \, ,
\ee
where $k = q/m$ denotes the specific charge. As a result, the equations are separated, and the trajectories are regular. Our result is consistent with \cite{Takahashi:2008zh}, where the authors used numerical methods to study the trajectories in the Kerr spacetime and found that almost all trajectories become regular in the limit $a \rightarrow M$. For convenience, we rewrite the equations of motion in the following forms
\bea
&& 2\G\f{dt}{d\tau} = \f{E+Lr}{r^2} \, , \label{ttau}\\
&& 2\G\f{d\phi}{d\tau} =\f{L}{\L^2} - \f{E+Lr}{r} - k \f{f(\t)}{\L^2} \, , \label{dphi} \\
&& 2\G\f{dr}{d\tau} = \pm_r \sqrt{R(r)} \, , \\
&& 2\G\f{d\t}{d\tau} = \pm_{\t} \sqrt{\T(\t)} \, . \label{dtheta}
\eea
Here the symbols $\pm_r$, $\pm_{\t}$ denote the sign of $U^r$ and $U^{\t}$, respectively, and we have introduced the radial and angular potentials
\bea
&& R(r) = -Cr^2 + 2ELr +E^2 \, , \label{R} \\
&& \T(\t) = C + L^2-2\G - \f{1}{\L^2}[L - k f(\t)]^2 \, .
\label{Theta}
\eea
The potentials must be non-negative along the trajectories. Note that only the angular potential Eq.~\eqref{Theta} gets changed by the presence of MSEM field, compared with the ones without electromagnetic field. This is one of the main results in this work. One can see that when the specific charge $k=0$, Eq.~\eqref{Theta} reduces to the form for geodesics in NHEK. However, since $f(\theta)$ is an undetermined function that depends on the choice of the potential $\mathcal{A}_{\text{MS}}$, we need to carefully analyze the angular potential based on different electromagnetic field configurations. Moreover, as $k$  increases, the conserved quantities $L$ and $C$ also increase, and the term “$2\G$'' in Eq.~\eqref{Theta} becomes negligible. Nevertheless, we will focus on the general case and retain the ``$2\G$'' term in our discussion.

 In addition, although the MSEM field also affects the motion of $\p$, the effect is somehow trivial. In fact,  from Eq.~\eqref{trans}, we learn that  $d\td \phi = d\phi + dt/\epsilon \sim dt/\epsilon$, which means $\phi$ always increases for the future-directed particles.

\subsection{The classification of radial motions}\label{sec32}
As the radial potential keeps invariant, the classification of the motions of charged particles in $r$-direction is the same as the ones of the geodesics in NHEK, which has been discussed in \cite{Kapec:2019hro, Compere:2020eat}. We give a brief review here. At first, since $t$ is the global time in NHEK, a future-directed trajectory should satisfy $dt/d\tau >0$ which gives
\be
r \gtrless r_c = -\f{E}{L} \, , \ \ \ \text{if} \ \ L \gtrless 0 \,,
\ee
from Eq.~\eqref{ttau}. Then, the radial potential in Eq. (\ref{R}) can be rewritten as 
\bea
R(r) = -C (r-r_-)(r-r_+),
\eea
with the roots $r_{\pm} = ELC^{-1}( 1 \pm \sqrt{1+C L^{-2}} )$. One can compare the values of $r_{\pm}$ with $r_c$ to find the allowed  regions of motion. We list the classification below and present the phase space in Fig.~\ref{fig:radial}.
\begin{figure}[h!]
	\centering
	\begin{tikzpicture}
	\node at (-6,2)
	{\includegraphics[scale=0.5]{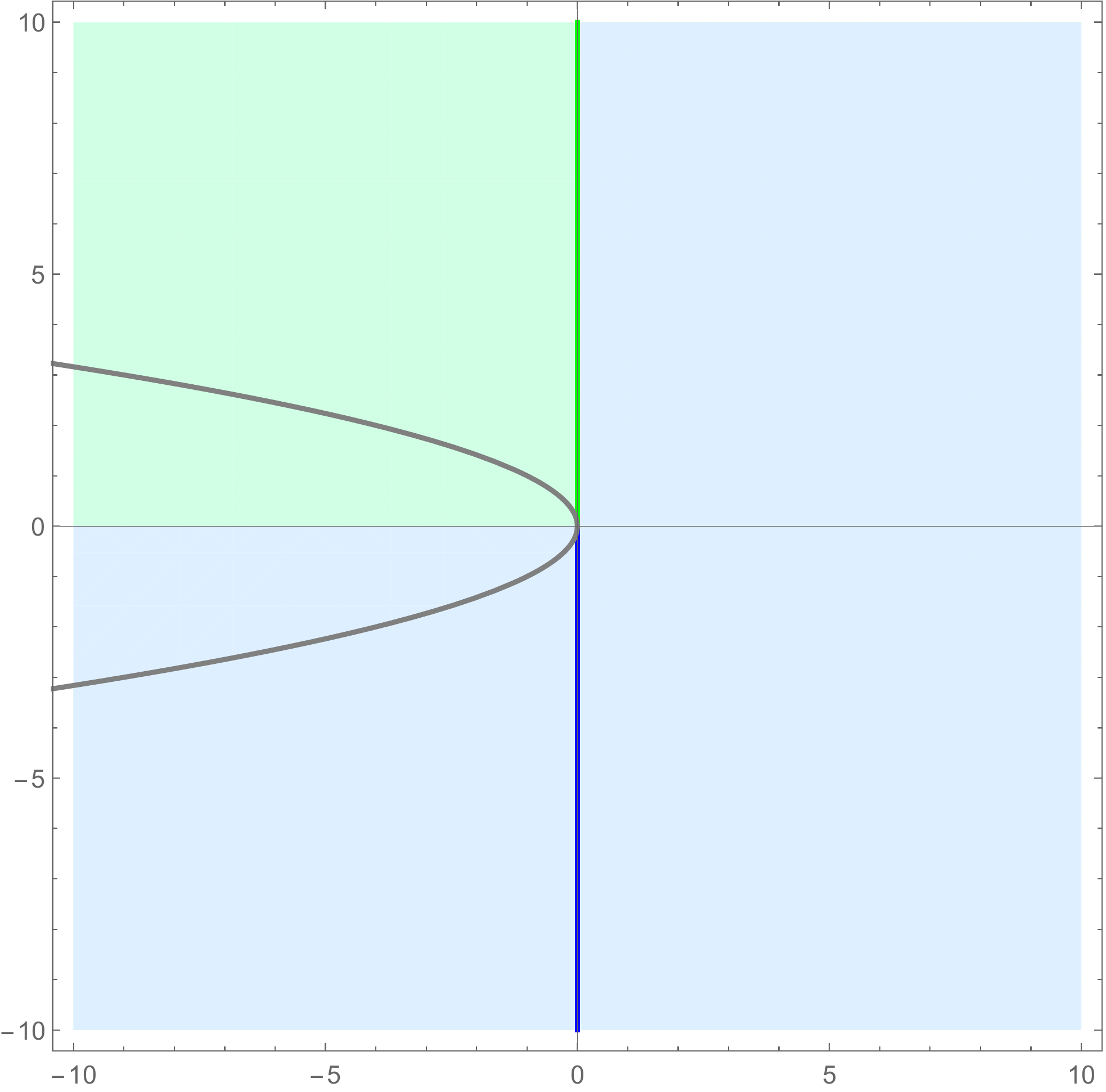}} ;
	\node[below left] at (-2,5) {\color{black} $E>0,\ [0,r_+]$};
	\node[below left] at (-2,0) {\color{black} $E>0,\ [0,r_-]$};
	\node[below left] at (-7,5.5) {\color{black} $E>0,\ [0,+\infty)$};
	\node[below left] at (-6.8,5) {\color{black} $E<0,\ [r_+,+\infty)$};
	\node[below left] at (-7,0) {\color{black} $E>0,\ [0,r_-]$};
	\node[below left] at (-7.4,4) {\color{gray} $C+L^2=0$};
	\node[below left] at (-7.5,3.2) {\small\color{black} $E>0,\ [0,+\infty)$};
	\node[below left] at (-7.4,2.7) {\small\color{black} $E<0,\ [r_c,+\infty)$};
	\node[below left] at (-7.9,2) {\small\color{black} $E>0,\ [0,r_c]$};
	\node[below left] at (-3.2,3.5) {\small\color{black} $E>0,\ [0,+\infty)$};
	\node[below left] at (-3.2,1.5) {\small\color{black} $E>0,\ [0,r_c/2]$};
	\node[below left] at (-5.5,-3) {\color{black} $C$};
	\node[below left] at (-10,3,2) {\color{black} $L$};
	\end{tikzpicture}
	\caption{Phase space of future-directed radial motions.}
	\label{fig:radial}
\end{figure}
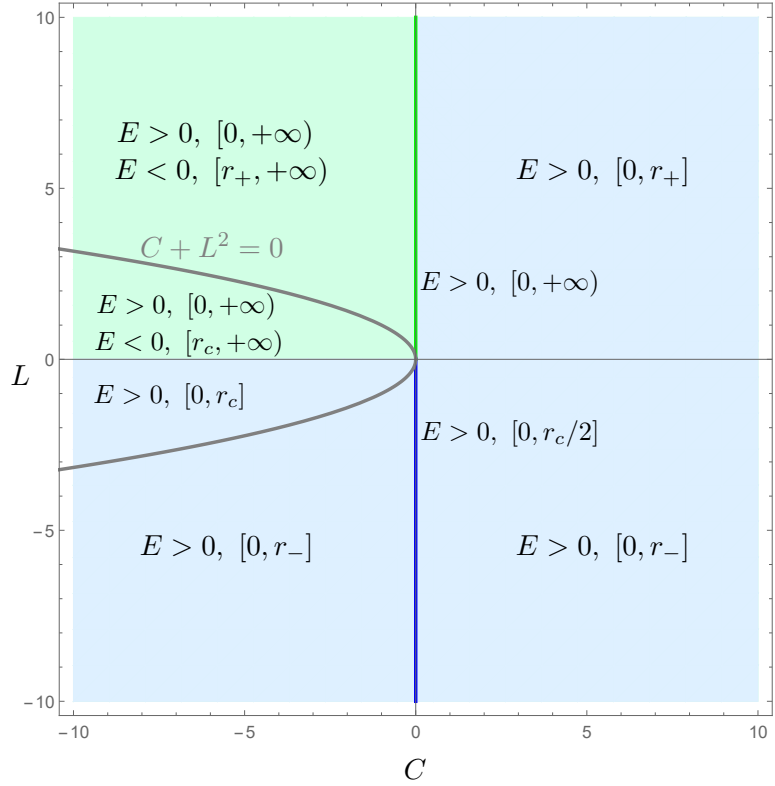
\begin{itemize}
\item Case 1: $C>0$ or $C > -L^2 \, , \, L<0$. In this case, $E>0$ must be satisfied. There is an outer bound at $r_{\pm_L}$, where $\pm_L$ is the sign of $L$. Thus the particles are confined in the region $0\le r\le r_{\pm_L}$ and can not escape to infinity. In Fig.~\ref{fig:radial}, this case is shown as the light blue region outside the gray curve.
\item Case 2: $-L^2<C<0 \, , \, L>0$. If $E>0$, the allowed region is $r\in [0,+\infty)$ with no turning point, while $E<0$, the allowed region becomes $r\in[r_+,+\infty)$, where $r_+$ is the turning point of the particles. In Fig.~\ref{fig:radial}, we use the light green region outside the gray curve to represent this case.
\item Case 3: $C<-L^2$. In this case, both $L, E <0$ is forbidden. If $L, E>0$, the allowed region is $r\in[0,+\infty)$. If $L>0,E<0$, the allowed region is $r\in[r_c,+\infty)$. If $L<0,E>0$, the allowed region is $r\in[0,r_c]$. For the latter two cases, the trajectories can reach $r_c$ and change their time directions. In Fig.~\ref{fig:radial}, this case is indicated by the region inside the gray curve. However, this case is excluded by the constraint from $\t$-motion \cite{Kapec:2019hro} for the geodesics of timelike particles. In the following, we can see that it is also excluded in both the force-free solution Eq.~\eqref{FF} and the Wald solution Eq.~\eqref{Wald}.
\item Case 4: $C = 0$. In this case, $E>0$ is required. If $L>0$, the allowed region is $[0,+\infty)$, which is represented by the green line in Fig.~\ref{fig:radial}. If $L<0$, the allowed region is $[0,r_c/2]$, represented by the blue line in Fig.~\ref{fig:radial}.
\end{itemize}

\section{The angular motions} \label{sec4}

In this section, we investigate the angular motions of charged particles in the NHEK geometry, in the presence of different MSEM fields. We consider two kinds of MSEM fields, one being the force-free one, the other being the vacuum solution. We pay our attention to not only the equatorial but also off-equatorial motions.

\subsection{General properties}\label{gen}
We start with the general properties of the angular potential Eq.~\eqref{Theta}. For an MSEM field, the electromagnetic invariant is
\bea
-F^2= -F_{\m\n}F^{\m\n} = \f{1}{2\G^2}\bigg[ f(\t) + \f{\partial_\theta f(\t)}{\L} \bigg] \bigg[ f(\t) - \f{\partial_\t f(\t)}{\L} \bigg] \,.
\eea
The field strength is electric-dominant when $-F^2>0$, and is magnetic-dominant when $-F^2>0$. To avoid the singularities at the poles, $\pl_\t f(0)$ and $\pl_\t f(\pi)$ should be zero. Near the north pole, the angular potential can be expanded as 
\bea
\T(\t) \approx C + L^2 - 2 - \bigg[ \f{k f(0)-L}{\t} \bigg]^2 + \mO(\t) \, .
\eea
Note that for the angular potential near the south pole, we only need to replace $f(0)$ by $f(\pi)$, so  we only give a discussion for the north pole. If $L \neq k f(0)$, the potential diverges at the pole such that particles can never reach the pole. If $L = k f(0)$, we have $\T(0) = C + k^2 f^2(0) - 2$ and $\T'(0)=0$. Thus, for $C\geq 2 - k^2 f^2(0)$, the particles can reach the north pole, and when the equality holds, the particles can move along the axis.

Secondly, the equatorial motions take place only when
\bea
\label{g1}
&&\T(\pi/2) =C+L^2-1-\f{1}{4}[k f(\pi/2)-L]^2 = 0 \,,  \\ 
&&\pl_\t\T(\pi/2)=-\f{k}{2}\pl_\t f(\pi/2)[k f(\pi/2)-L]= 0 \, .\label{g2}
\eea
When $\pl_\t f(\pi/2) = 0$,  Eq. \eqref{g2} is always true so that the equatorial motions requires  the condition Eq.~\eqref{g1}, which gives a curve in the parameter space of $(C, L)$. But when $\pl_\t f(\pi/2) \neq 0$, there must be $L = k f(\pi/2)$  for the equatorial motions. As a result, in the parameter space the equatorial motions correspond to a point $(1-L^2, L=k f(\pi/2))$ combined with the solution curve of Eq.~\eqref{g1}.  Moreover, the stability of equatorial motion is determined by
\bea\label{sc}
\pl^2_\t\T(\pi/2) = -2 - \f{1}{2}[k \pl_\t f(\pi/2)]^2-\f{1}{2}[L-k f(\pi/2)][3L-3k f(\pi/2)-k \pl^2_\t f(\pi/2)] \, .
\eea
The unstable/stable motions correspond to $\pl^2_\t\T(\pi/2) \gtrless 0$. If $\pl_\t f(\pi/2) = 0$, $\T''(\pi/2)$ becomes a function of $L$, so that the stability depends on the value of $L$. In the case $\pl_\t f(\pi/2) \neq 0, L=k f(\pi/2)$,  the equatorial trajectories are always stable since the term $-2 - \f{1}{2}[k \pl_\t f(\pi/2)]^2$ in the Eq. \eqref{sc} is always negative.

\subsection{Force-free solution}\label{FFno}
\begin{figure}[h!]
	\centering
	\begin{tikzpicture}
	\node at (-6,2)
	{\includegraphics[scale=0.43]{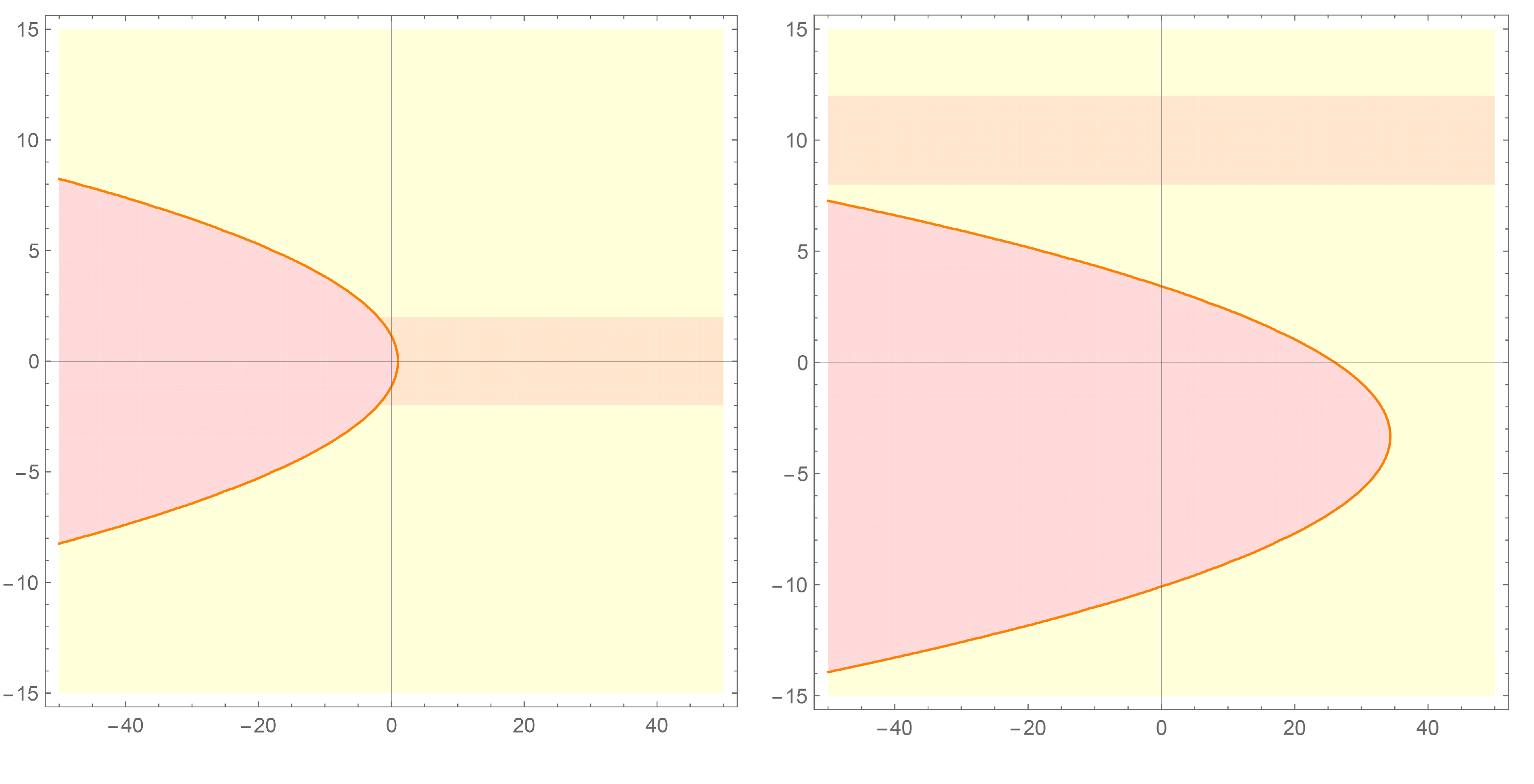}};
	\node[below left] at (-9,-1.4) {\color{black} $C$};
	\node[below left] at (-1.1,-1.4) {\color{black} $C$};
	\node[below left] at (-13.5,2.2) {\color{black} $L$};
	\node[below left] at (-5.6,2.2) {\color{black} $L$};
	\node[below left] at (-12,5.7) {\small\color{black} $\mE = 0$};
	\node[below left] at (-4,5.7) {\small\color{black} $\mE = 10$};
	\node[below left] at (-10.5,2.3) {\large\color{red} Forbidden};
	\node[below left] at (-1.5,2.3) {\large\color{red} Forbidden};
	\node[below left] at (1,0) {\large\color{black} Oscillatory};
	\node[below left] at (-7,0) {\large\color{black} Oscillatory};
	\node[below left] at (-7,3.9) {\color{black} $[0,z_+]$};
	\node[below left] at (1,3.9) {\color{black} $[0,z_+]$};
	\node[below left] at (-7,2.4) {\color{black} $[0,z_-]$};
	\node[below left] at (1,4.9) {\color{black} $[0,z_-]$};
	\end{tikzpicture}
	\caption{Phase spaces of angular motions in the force-free field.}
	\label{fig:ff}
\end{figure}
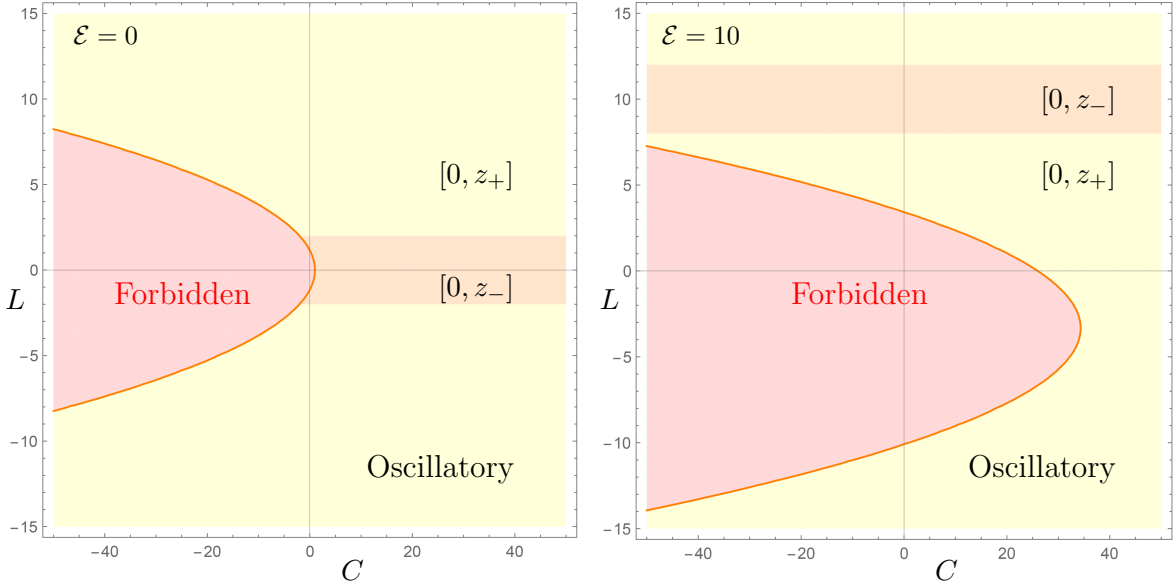

In this subsection, we focus on the force-free solution\footnote{The force-free fields have been explored to approximate black hole magnetospheres \cite{Compere:2015pja}. In the NHEK geometry, the force-free condition is equal to the degenerate condition, $(\star F)_{\m\n}F^{\m\n} = 0$.} in which the source feels a vanishing Lorentz force, 
\be
\label{ffe}
J^{\m}F_{\m\n} = 0 \,.
\ee
On the other hand, from the Maxwell equation $\nabla_{\m}F^{\m\n} = J^{\n}$, we can get
\bea
J^{\m}\pl_{\m}= -\f{1}{2\G^2\L}\bigg[ \pl_{\t}\bigg(\f{\pl_{\t}f(\t)}{\L}\bigg)+\L f(\t) \bigg]\pl_{\phi} \, ,
\label{J00}
\eea
which is a pure toroidal current since only the $\phi$-component of $J^\mu$ survives. Combining with Eq. \eqref{ffe}, we find that $f(\t) = f_0$ must be a constant and
\bea
\mA= f_0(d\phi+rdt) \, , \quad J^{\m}\pl_{\m} = -\f{2 f_0}{1+\cos^2{\t}}\pl_{\phi} \,,
\label{FF}
\eea
which describes an electric-dominated field outside the ergosphere\footnote{There is no global timelike Killing vector in NHEK, and only the region outside the ergosphere are the physical region of electromagnetic fields \cite{Kay:1988mu}.}. Considering the fact that the system of interest has the $Z_2$ symmetry ($\t\rightarrow \pi-\t$), so we only focus on $0 \leq \t \leq \pi/2$ without losing generality. It is convenient to define
\bea
z=\cos^2{\t}\,,
\eea
then the Eq.~\eqref{dtheta} and Eq.~\eqref{Theta} can be rewritten as
\be\label{dtFF}
\G\f{dz}{d\tau} = \pm_{\t} \sqrt{z \T_{\text{FF}}(z)} = \pm_{\t} \sqrt{z(\ma z^2 + \mb z +\mc) } \, ,
\ee
\bea
&&\ma = 1-\f{1}{4}(L-\mE)^2 \, , \ \
\mb = -C - \f{3}{2}L^2 +\mE L - \f{1}{2} \mE^2 \, , \nn \\
&&\mc = C +L^2 -1 - \f{1}{4} (L-\mE)^2
\label{abcFF}
\eea
with $\mE = k f_0$ being the electromagnetic parameter. The angular potential  $\T_{\text{FF}}(z)$  has two roots $z_{\pm} = \f{1}{2\ma}(-\mb\pm\sqrt{\mb^2-4\ma\mc})$. Note that at the north pole, $z=1$, and the angular potential $\T_{\text{FF}}(z)$ becomes
\be
\T_{\text{FF}}(1)=\ma + \mb + \mc =-(L-\mE)^2 \, ,
\label{T11}
\ee
which is negative when $L\neq\mE$. Considered $\T_{\text{FF}}(0)=\mc$,  the allowed region of motion is $z\in[0, z_{\mp_{\ma}}]$ with $\mp_{\ma}$ being minus the sign of $\ma$, when $\mc>0$. In this case, the particles oscillate between $\t_- =\cos^{-1}{\sqrt{z_{\mp_{\ma}}}}$ and $\t_+ =\pi-\cos^{-1}{\sqrt{z_{\mp_{\ma}}}}$, crossing the equatorial plane for each oscillation. These trajectories are usually called ``oscillatory'' motions. The yellow and orange regions in the phase spaces in Fig. \ref{fig:ff} represent the oscillatory motions for $\ma<0$ and $\ma>0$, respectively. 
Actually, in the force-free field, there holds 
\be
\mc = C+L^2-1 - \f{1}{4} (L-\mE)^2 \geq C + L^2 -2\G- \f{1}{\L^2} (L-\mE)^2 = \T_\text{FF}(z) \geq 0 \, .
\label{cgeq0}
\ee
 In Fig. \ref{fig:ff} the case $\mc<0$ is indicated by the red color, labeled as the forbidden regions. When $\mc = 0$, the particles can move in the equatorial plane with
\be
C_{\text{eq}} = 1 - \f{3}{4}L_{\text{eq}}^2-\f{1}{2}\mE L_{\text{eq}} + \f{1}{4} \mE^2 \, .
\ee
And, the stability is determined by $ \partial_z\T_{\text{FF}}(0) = -1 -3(L_{\text{eq}}-\mE)^2/4 < 0 $. Thus, all the equatorial motions are stable, represented by the orange curves in Fig. \ref{fig:ff}.

Except for the special case $\mc=0$, we can conclude that only the oscillatory motions are allowed. This picture can be inferred from the study of the geodesics in the NHEK geometry. It has been shown in \cite{Kapec:2019hro} that the NHEK geodesics are all oscillatory and the angular potential of the NHEK geodesics takes the form of Eq. \eqref{Theta} with $f(\theta)=0$. In our force-free case, we have $f(\theta)=f_0$ as a constant so that the angular potential is simply modified by adding a constant.

 Moreover, it can be checked that $ C+L^2 > \mc \geq 0$ for all the equatorial and off-equatorial motions. This ensures that a charged particle can never reaches $r_c$ and the time direction is kept unchanged in the force-free field considering the radial potential discussed in Sec. \ref{sec32}.

\subsection{Vacuum solution}\label{V}

Next we turn to another class of MSEM fields, which are produced by the sources outside the NHEK region. In this case, we have $ J^{\m} = 0$ in the Eq.~\eqref{J00} and obtain the vacuum solution
\begin{align}
\label{vac}
\mA=& \, \bigg( A_M \f{\sin^2{\t}}{1+\cos^2{\t}} + A_E \f{2\cos{\t}}{1+\cos^2{\t}} \bigg)(d\p + rdt).
\end{align}
The parameters $A_M, A_E$ are real constants, which can be taken as magnetic and electric fields in the sense that near the axis, the electromagnetic invariant becomes 
\bea
-F^2 \approx (A_E^2-A_M^2)/2\,.
\eea
We define
\bea
u = \cos{\t}\,,
\eea
and the angular equation can be rewritten as
\be
2\G\f{du}{d\tau} = \mp_{\t} \sqrt{\T_{\text{v}}(u)} \, ,
\ee
with the angular potential
\begin{align}
\label{Tvac}
\T_{\text{v}}(u) =& \bigg(1-\f{L^2}{4}\bigg)u^4-\bigg(C+\f{3}{2}L^2\bigg)u^2+C+\f{3}{4}L^2-1 \nn \\
&- \bigg( \f{k A_M L}{2}+\f{k^2 A^2_M}{4} \bigg)u^4 - (k A_E L+k^2 A_M A_E)u^3 +k^2\bigg( \f{A^2_M}{2}-A^2_E \bigg)u^2 \nn \\
&+ (k A_E L -k^2 A_M A_E )u +k\f{ A_M L}{2} -k^2\f{A^2_M}{4} \, .
\end{align}
One can see that this potential is invariant under $L \rightarrow -L$, $A_M \rightarrow -A_M$ , $A_E \rightarrow -A_E$. Note that, the system does not have $Z_2$ symmetry ($\t\rightarrow \pi-\t$), which complicates the classification of motions.

We consider two special cases here, before we discuss the more general cases. The first special case is that the equatorial motions, which requires 
\bea
\T_{\text{v}}(0) = \pl_u\T_{\text{v}}(0) = 0\,,
\eea
and then
\bea\label{leq}
&& L_{\text{eq}} =
\begin{cases}
	\, \text{arbitrary value},& \quad \text{if}\,\,\, \ A_E = 0 \, ; \nn \\
	\, k A_M,& \quad \text{if}\,\,\, \ A_E \neq 0 \, . \nn
\end{cases}
\\
&& C_{\text{eq}} = 1 + \f{1}{4}( k^2A_M^2 -2 k A_M L_{\text{eq}}-3L_{\text{eq}}^2 ) \, .
\label{eqv}
\eea
As discussed in Sec. \ref{gen},  if $A_E = 0$, there exists a class of equatorial motions with various $L$, but if $A_E \neq 0$, the equatorial trajectories must have fixed parameters. The stability is determined by the sign of
\be
\pl^2_u\T_{\text{v}}(u) |_{\text{eq}} =-1-k^2A_E^2 +\f{1}{4} (k A_M + 3L_{\text{eq}})(k A_M -L_{\text{eq}}) \,,
\ee
where the subscript `` eq '' means that the function takes value at $L=L_{\text{eq}}, C=C_\text{eq}$ and $u=0$. For a non-vanishing $A_E$, $\pl^2_u\T_{\text{v}}(u) |_{\text{eq}}=-1-k^2A_E^2 < 0$, which means that  the equatorial trajectories are always stable. The case $A_E = 0$ with generic trajectories will be discussed carefully  in the next subsection.

The second special case is a type of motion that lies on conical surfaces which are determined by
\bea \label{conicalv}
\T_{\text{v}}(u) = 0 \, , \ \ \ \pl_u \T_{\text{v}}(u) = 0 \, , \ \ u \neq 0\, .
\eea
Such kind of conical motion is absent in the NHEK geodesics. In this kind of motion, 
the conserved quantities in the parameter space are constrained by the above relation \eqref{conicalv},
\bea
&& L_{\pm}(u) = \f{ k A_M (1-u^2)^2u+k A_E(u^4-4u^2-1) \pm (1-u^2)X}{u(3-u^2)(1+u^2)} \, , \nn \\
&& C_{\pm}(u) = \f{Y \mp\big[2k A_M(3-2u^2+3u^4)u-k A_E(3+15u^2-3u^4+u^6)\big](1-u^2)X}{2(3-u^2)^2(1+u^2)^2u^2} \, ,\nn \\
\l. \r.
\label{pmv}
\eea
where
\begin{align}
X =&\sqrt{ k^2 A_M^2(u^2+4)u^2+k^2 A_E^2(1-u^2)^2+4k^2 A_MA_E(1-u^2)u-4(3-u^2)(1+u^2)u^2} \, ,\nn \\
Y =&
-2k^2A_M^2\big(3-14u^2+12u^4-2u^6+u^8\big)u^2
-k^2A_E^2\big(3+9u^2+42u^4-26u^6+3u^8+u^{10}\big) \nn \\ &+ 32k^2A_MA_E(1-u^2)^2u^3+4\big(9+2u^4+8u^6-3u^8\big)u^2 \, .
\label{XYv}
\end{align}
Note that $X^2$ must be non-negative, which gives a constraint on the allowed region of $u$. As these expressions are very complicated, we would not like to give a detailed discussion on this case. Instead, we will focus on a simple case $A_E=0$ in the following, since $A_E=0$ implies that the gauge potential is magnetic-dominant which is a more realistic situation in astrophysics.

\subsubsection{Angular potential and off-equatorial motions}

In the following, we take $A_E=0$ in the vacuum solution.  
When $A_E=0$, the vacuum solution Eq.~\eqref{vac} actually reduces to the Wald potential in the near-horizon limit\cite{Wald:1974np}. To clarify this point, recall that the Wald vector potential in the BL coordinates takes the form
\bea
\label{W1}
\mA_{\text{W}} = (aB-Q/2)\pl_{\td{t}} + (B/2) \pl_{\td{\p}} \, ,
\eea
where $Q$ is the charge of the black hole, and $B$ is a test magnetic field vertical at infinity. When zooming into the NHEK region by Eq.~\eqref{trans}, we have 
\bea
2\pl_{\td{t}} = \ep^p\pl_t - \pl_{\p} \, , \, \pl_{\td{\p}}=\pl_{\p}\,.
\eea
In the leading order,  only the field produced by the charge survives, and the vector potential reduces to\cite{AlZahrani:2010qb}
\be\label{Wald}
\mathcal{A}_{\text{W}} = \f{Q}{4} \pl_{\phi} \, ,
\ee
which is the same as the field \eqref{vac} if we identify $Q=A_M$. The physical reason why we drop the test magnetic field in the Wald vector potential is that the extreme black hole behaves like a perfect diamagnet, excluding the lines of test magnetic fields \cite{Chamblin:1998qm}, and only the intrinsic field provided by $Q$ survives. Moreover, although the test magnetic term in the potential is dropped, the charge $Q$ still generates a magnetic dipole due to the fast rotation of the black hole, so that the Eq.~\eqref{Wald} describes a magnetic-dominated case, i.e., $-F^2<0$ outside the ergosphere. 

Now we turn to study the angular potential in the background of  Eq.~\eqref{Wald}. We will concern  the off-equatorial motions especially. Note that the system has an $Z_2$ symmetry ($\t\rightarrow \pi-\t$), so we consider $0 \le \t \le \pi/2$ without loss of generality. By defining 
\bea
z=\cos^2{\t}\,,
\eea
we can obtain a quadratic angular potential and find 
\be\label{polar0}
\G\f{dz}{d\tau} = \pm_{\t} \sqrt{z \T_{\text{W}}(z)} = \pm_{\t} \sqrt{z(\ma z^2 + \mb z +\mc) } \, ,
\ee
\bea
&& \ma = 1-\f{1}{4}(L+\mB)^2 \, , \, \mb = -C - \f{3}{2}L^2 + \f{1}{2}\mB^2 \, , \nn \\
&& \mc = C + \f{1}{4}(L + \mB)(3L - \mB)-1 \, ,
\label{abc}
\eea
where we have introduced an electromagnetic parameter $ \mB = k Q $. In addition, considering that the potential is invariant under $L \rightarrow -L$, $\mB \rightarrow -\mB$,  we may set $\mB \geq 0$ for simplicity. It should be stressed that the Eq.~\eqref{polar0} has appeared in \cite{AlZahrani:2010qb}, where the authors investigated the special cases that charged particles move along the axis and on the equatorial plane. However, in the present work we mainly focus our attention to  general off-equatorial motions and quantitatively study the effect of the electromagnetic field. Hence, we need a detailed analysis of the angular potential in the Eq. \eqref{polar0}.

Obviously, the equation 
\bea
\T_{\text{W}}(z)=\ma z^2 + \mb z +\mc=0
\eea
has two roots at $z_{\pm} = (-\mb\pm\sqrt{\mb^2-4\ma\mc})/2\ma$. And at the poles the potential becomes 
\be
\T_{\text{W}}(1)=\ma + \mb + \mc =-L^2 \, .
\label{T1}
\ee
The special case $L=0$ indicates that the trajectories might cross the pole. It is beyond our interest, so we focus on $L\neq0$ in the following. Therefore, we have $\T_{\text{W}}(1)<0$. On the other hand, we find  $\T_{\text{W}}(0)=\mc$. It is worth mentioning that for the force-free case, the angular potential is always non-negative at $z=0$, considering Eq.~\eqref{cgeq0}. However, due to a nontrivial influence of $\mB$ on the angular potential in the vacuum solution, the sign of $\mc$ is uncertain, and the motions of charged particles need a detailed discussion.

\begin{figure}[h!]
	\centering
	{\includegraphics[scale=1.2]{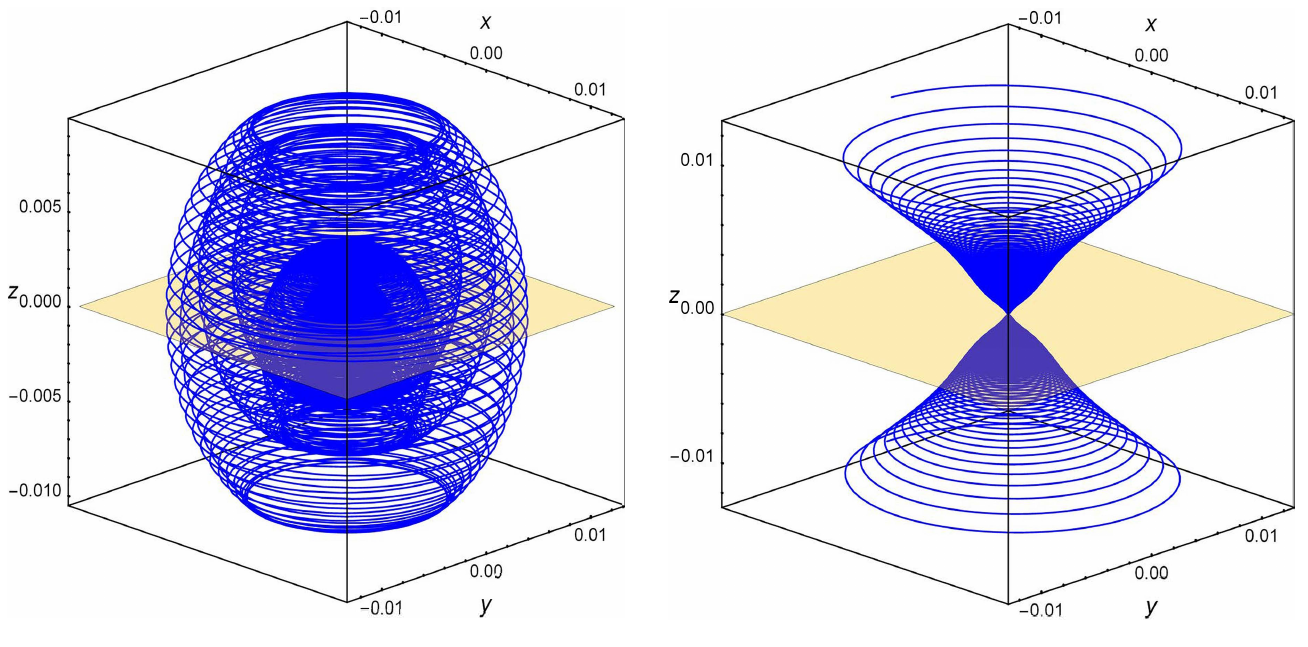}}
	\caption{Left: The trajectory of an oscillatory motion with the initial parameter chosen as $\mB=50$, $(E, L, C) = (20, 30, 100)$, $r=0.01$ and $\t = \pi/2$. Right: The trajectories of two symmetric vortical motions with the initial parameter chosen as $\mB=50$, $(E, L, C) = (5, 15, -220)$, $r = 0.01$ and $\t =\pi/4 (\text{above}), 3\pi/4(\text{below})$. For a better display, the trajectories are plotted in the coordinate: $x = \ep_0 r \sin{\t}\cos{(\p + t/\ep_0)}$, $y = \ep_0 r \sin{\t}\sin{(\p + t/\ep_0)}$, $z =\ep_0 r \cos{\t}$, with $\ep_0 = 0.001$. The event horizon is at $(0,0,0)$, and the yellow planes denote the equatorial plane.}
	\label{fig:Tra}
\end{figure}

\begin{figure}[h!]
	\centering
	\begin{tikzpicture}
	\node at (-15,2)
	{\includegraphics[scale=0.66]{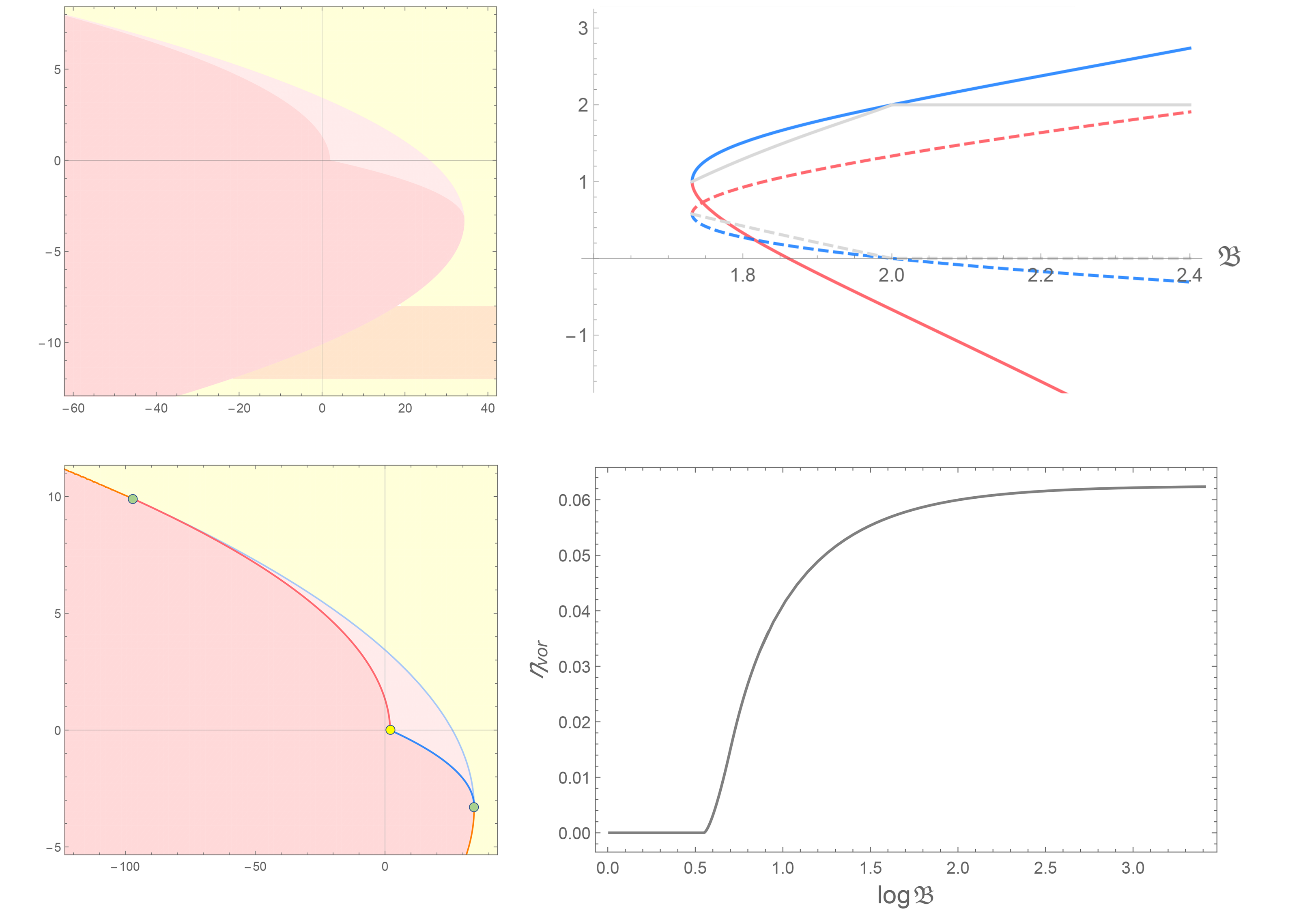}};
	\node[below left] at (-21.5,3.5) {\color{black} $\mB=10$};
	\node[below left] at (-21.5,-2.9) {\color{black} $\mB=10$};
	\node[below left] at (-18.5,2.6) {\color{black} $C$};
	\node[below left] at (-23.5,5.5) {\color{black} $L$};
	\node[below left] at (-20,5.6) {\large\color{red} Forbidden};
	\node[below left] at (-17.3,7.9) {\large\color{black} Oscillatory};
	\node[below left] at (-17.9,6.4) {\small\color{black} Vortical};
	\node[below left] at (-7,8.2) {\small\color{purple} $C_-^0$};
	\node[below left] at (-7.8,7.3) {\small\color{gray} $C_m$};
	\node[below left] at (-8.6,3.4) {\small\color{red} $C_+^0$};
	\node[below left] at (-7.8,5.3) {\small\color{gray} $L_m$};
	\node[below left] at (-7,4.4) {\small\color{purple} $L^0_-$};
	\node[below left] at (-8.6,6.5) {\small\color{red} $L^0_+$};
	\node[below left] at (-19.3,-0.4) {\small\color{purple} $+$};
	\node[below left] at (-17.8,-2.1) {\small\color{red} $-$};
	\node[below left] at (-17.5,-2.6) {\small\color{purple} $(C^0_-,L^0_-)$};
	\node[below left] at (-18.7,-1.7) {\small\color{gray} $(C_m,L_m)$};
	\node[below left] at (-21.5,1.2) {\small\color{red} $(C^0_+,L^0_+)$};
	\node[below left] at (-18.5,-3.8) {\color{black} $C$};
	\node[below left] at (-23.5,-0.8) {\color{black} $L$};
	\end{tikzpicture}
	\caption{Top left: Phase space of angular motions in the Wald potential with $\mB = 10$. The phase spaces with larger $\mB$ are qualitatively similar.  Down left: Endpoints of two branches of conical motions with $\mB = 10$. Top right: The endpoints and the cusp are plotted as the functions of $\mB$. The plot region is $\mB \in [1.6, 2.4]$. Bottom right: $\eta_{\text{vor}}$ as a function of $\log{\mB}$. The plot region is $\mB \in [1, 30]$.}
	\label{fig:vor}
\end{figure}
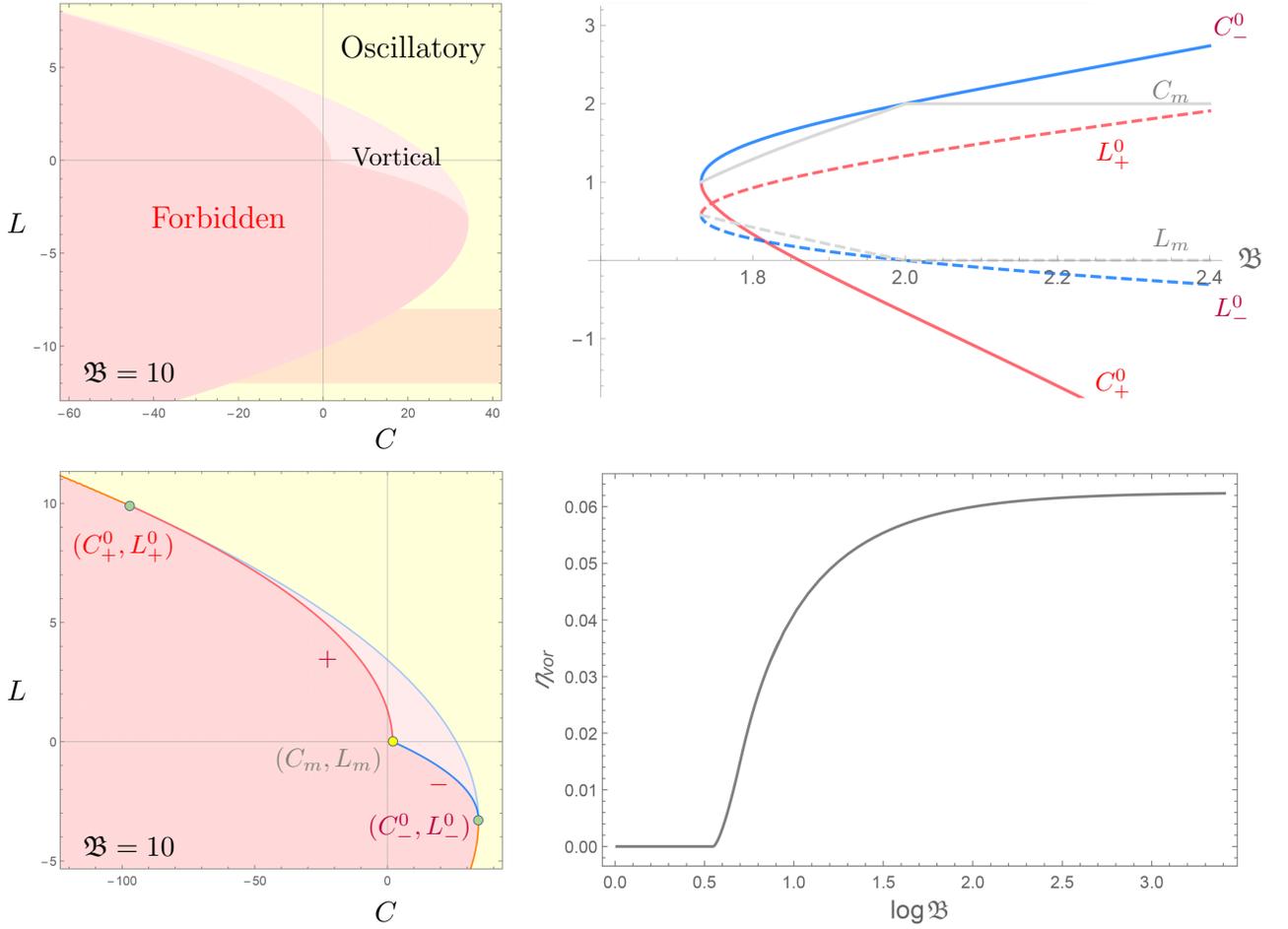

We first consider the case $\mc>0$. In this case $\T_{\text{W}}(0)>0$ so that the charged particles move along normal oscillatory trajectories, and the allowed region is $z\in[0, z_{\mp_{\ma}}]$, with $\mp_{\ma}$ being minus the sign of $\ma$. As an example,  an oscillatory trajectory is shown in the left plot in Fig. \ref{fig:Tra}. We also present the oscillatory motions for $\ma<0$ and $\ma>0$ in the phase space indicated by the yellow and orange regions in the top left figure of the Fig. \ref{fig:vor}, respectively, where we set $\mB=10$.


Then we move to the case $\mc<0$. The motions of charged particles must satisfy $z\in[0,1]$, such that the coefficients have to obey
\bea\label{abc}
\ma < 0 \, , \ \ 0<-\f{\mb}{2\ma} <1 \, , \ \ \mb^2-4\ma\mc \geq 0 \, .
\eea
If the above inequalities can be satisfied, the polar angles of charged particles oscillate between $\theta_-=\cos^{-1}{\sqrt{z_{+}}}$ and $\theta_+=\cos^{-1}{\sqrt{z_{-}}}$ , or $\pi-\theta_+$ and $\pi-\theta_-$ , without crossing the equatorial plane, and such trajectories are the so-called ``vortical'' motions. Likewise, we present an example of a pair of vortical motions in the right plot of Fig. \ref{fig:Tra}, which are symmetric about the equatorial plane and reflects the symmetric transformation $\t\rightarrow\pi-\t$, $\p\rightarrow \p+\pi$, $\pm_{\t}\rightarrow \mp_{\t}$ in Eq.~\eqref{polar0}. In addition, in the top left figure in Fig.~\ref{fig:vor}, we also show the vortical motions in the phase space which are indicated by the grey color. Considering the fact that there are no vortical motions for the geodesics in the NHEK region \cite{Kapec:2019hro}, we can easily conclude that these vortical motions are caused by the electromagnetic force.  

At last, we give a brief discussion on the case $\mc=0$. In this case, we have $\T_{\text{W}}(z=0)=\mc=0$ which corresponds to 
\bea
\T_{\text{W}}(\theta=\frac{\pi}{2})=\partial_\t \T_{\text{W}}(\theta=\frac{\pi}{2})=0\,,
\eea
and 
\bea
\partial^2_\theta\T_{\text{W}}(\theta=\frac{\pi}{2})\propto\partial_z\T_{\text{W}}(z=0) = -1 + (\mB+3L)(\mB-L)/4\,.
\eea 
When $\partial_z\T_{\text{W}}(z=0)\le0$, the potential  $\T_{\text{W}}(z)$ is a decreasing function in $z\in[0, 1]$, thus the particles are confined to the equatorial plane and the trajectories are stable. When $\partial_z\T_{\text{W}}(z=0)>0$, $\T_{\text{W}}(z)$ increases to a peak and then decreases in $z\in[0, 1]$, and $z=0$ is an unstable extremal point of $\T_{\text{W}}(z)$, corresponding to unstable equatorial motion.

In addition, we want to emphasize that all the motions discussed above satisfy $ C+L^2 \geq 0$ (see proof in Appendix.~\ref{proof}), which means a charged particle never reaches $r_c$ and always keeps future-directed in  the Wald potential.

\subsubsection{More on vortical motions and astrophysical implications}\label{Conical}

In this subsection, we would like to illustrate the features of the phase space more carefully. In particular, we want to give a more detailed analysis of the vortical motions.

Firstly, it is worth mentioning that there are  vortical motions only in the vacuum solution, and no such motions in the force-free model. It is expected that the force-free model would be a good approximation of the magnetosphere away from the equatorial plane based on simulations. However, it should be noted that the force-free MSEM considered in this work assumes maximal symmetry, which is an oversimplification and idealization compared to normal force-free solutions in astrophysical research. It would be interesting to see if vortical motion can appear with the assumption being relaxed.
	
Then, let us focus on the boundaries of the vortical motions in the phase space. It is convenient to zoom in on the grey region of the top left plot in Fig. \ref{fig:vor} to get the bottom left plot. In the bottom left plot of Fig.~\ref{fig:vor}, one can see that the orange and light blue curves represent the stable and unstable equatorial motions at $\mc=0$. In other words,  the stable equatorial motions form the boundary of the oscillatory motions, and the unstable equatorial motions form the dividing line between the oscillatory and vortical motions in the phase space. Apparently, there could be other critical motions that form the boundaries of the vortical motions in the phase space.

For these additional critical motions, we have $z_- = z_+$ corresponding to a constant $\theta$, which is a simplified case of Eq.~\eqref{conicalv}. In this case, the charged particles are moving on conical surfaces and the trajectories are determined by
\bea
\T_{\text{W}}(z) = \T_{\text{W}}'(z) = 0\,.
\eea
For simplicity, we would like to call them the conical motions which are short for the motions on the conical surfaces. The corresponding conserved quantities take the form
\bea
L_{\pm}(z) = \f{\mB(z-1)^2\pm 2(1-z)X}{(z+1)(3-z)} \, , \ \ C_{\pm}(z) = \f{Y \mp 2\mB(3z^2-2z+3)(1-z)X}{(z+1)^2(3-z)^2} \, ,
\label{pm}
\eea
where we introduced
\bea
&& X = \sqrt{z^2-2z+\mB^2-3},\nn \\
&& Y = -(\mB^2+6)z^4 + 2(\mB^2+8)z^3 + 4(1-3\mB^2)z^2 + 14\mB^2 z + 18-3\mB^2\,.
\label{XY}
\eea
The condition  $X^2\ge0$ leads to 
\bea
z^2-2z+\mB^2-3\ge0\,, 
\eea 
which requires that $\mB^2-3$ has to be non-negative in order that the above inequality can be obeyed for $z\in[0, 1]$. As a result, we obtain a critical value of the electromagnetic parameter as 
\be
\mB_c = \sqrt{3}\,.
\ee
When $\mB<\mB_c$, there is no conical motion, and consequently the vortical motions do not exist as well. Simply speaking, $\mB\ge\mB_c$ is a necessary condition to have vortical motions.

When the condition $\mB\geq\mB_c$ holds, there are two branches $L_+(C_+)$ and $L_-(C_-)$ from Eq.~\eqref{pm}, corresponding to two smooth curves in the phase space. In the bottom left plot of Fig.~\ref{fig:vor}, we set $\mB=10>\sqrt{3}$ so that the vortical and conical motions exist; $L_+(C_+)$ and $L_-(C_-)$ are marked by the red and dark blue curves respectively,  and their intersection is given by $(C_m, L_m)$. Moreover, we use $(C_+^0, L_+^0)$ to denote the intersection of the light blue and the red curves, and $(C_-^0, L_-^0)$ to represent the intersection of the light blue and the dark blue curves, where the light blue one is a portion of the curve determined by $\mc=0$. Now, we see that when $\mB>\mB_c$ the existing vortical motions are bounded by the curves $L_+(C_+)$, $L_-(C_-)$ and $\mc=0$ with a few intersection points. 

The values of the intersection points $(C_+^0, L_+^0)$, $(C_-^0, L_-^0)$ and $(C_m, L_m)$  can be determined for a fixed $\mB$.  Combining $\mc=0$ and $L=L_+(C_+)$, we obtain that at $z=0$ 
\bea
\label{CLpm}
L_{\pm}(0) = L_{\pm}^0 = \f{1}{3}\mB \pm \f{2}{3}\sqrt{\mB^2-\mB_c^2} \, , \, \ C_{\pm}(0) = C_{\pm}^0 = 2- \mB L_{\pm}^0,
\eea
which means the solid angle of the conical plane is $2\pi$ and the conical plane coincides with the equatorial plane. On the other hand, the non-negativity of $X^2$ gives an upper limit of $z$, $z = z_m$. For the case $\mB_c <\mB<2$, we have $z_m = 1-\sqrt{4-\mB^2}$ and
\bea
L_{\pm}(z_m) = L_m = \f{4}{\mB}-\mB \, , \, \
C_{\pm}(z_m) = C_m = \sqrt{4-\mB^2} \bigg( \f{8}{\mB^2}-2 \bigg)-\mB^2 - \f{32}{\mB^2} + 14 \, .
\eea
For the case $\mB > 2$, we always have $z_m=1$, and $(C_m,L_m) \equiv (2,0)$, which corresponds to the motions along the spin axis. The behaviors of $C^0_{\pm}, L^0_{\pm}, C_m, L_m$ as the functions of $\mB$ are shown in the top right plot of Fig.~\ref{fig:vor}. For $\mB = \mB_c$, all the points coincide with $L = 1/\sqrt{3}$, $C = 1$. Moreover, in the case that $\mB_c<\mB<2$  all the conical motions are prograde as $\L_\pm^0>0$, while in the case that $\mB>2$, the conical motion could be prograde with $L_+^0 > 0$ as well as retrograde with $L_-^0<0$.

Next, we consider the stability of the conical motions. From Eq.~\eqref{polar0} and Eq.~\eqref{pm}, we can easily have $\partial_z^2\T_{\text{W}}(z) \leq 0$ which means the conical motions are always stable. Moreover, $L^0_+(C^0_+)$, $L^0_- (C^0_-)$ correspond to marginal stable motions, since in this case $\partial_z^2\T_{\text{W}}(z) = 0$.  Since the conical motions form the boundary of the vortical motions, they will change into the vortical motions and swing slightly around the original conical surface under a perturbation of the conserved quantities.


Furthermore, since the vortical motions are bounded, the area of the vortical motions in the phase space should be finite. To characterize the ratio of the vortical motions in the phase space, we would like to introduce a parameter $\eta_{\text{vor}}$ as
\be
\label{Sv}
\eta_{\text{vor}} = \int \limits_{\text{vor}}\f{dC dL}{S_0} \, ,
\ee
where ``vor'' means an integration in the vortical region, and $S_0$ is defined as the area of the rectangle\footnote{By Eq.~\eqref{pm} one finds that both $L_+(C_+)$ and $L_-(C_-)$ are decreasing functions of $C_+$ and $C_-$, respectively. Then in the phase space, the parabolic curve determined by $\mc = 0$ has an extreme point with $L=-\mB/3 < L^0_-$. Thus, the rectangle encloses the vortical region in the phase space. } bounded by the straight lines $C=C_\pm$ and $L=L_\pm$.  From Eq.~\eqref{CLpm} we have 
\bea
S_0 =(C^0_--C^0_+)(L^0_+-L^0_-) = 16\mB(\mB^2-3)/9\,. 
\eea
Strictly speaking,  $\eta_{\text{vor}}$ is not the ratio of the vortical motions in the whole phase space, it only characterize the ratio of the vortical motions in the space enclosed by the rectangle.
It is straightforward to compute $\eta_{\text{vor}}$,  with the help of the expressions of the conical motions. The bottom right plot of Fig.~\ref{fig:vor} shows the result of $\eta_{\text{vor}}$ as a function\footnote{When we extend to the regime $\mB < \mB_c$, the expression of $S_0$ still holds, but there is no  vortical motion. Nevertheless, we may define $\eta_{\text{vor}}\equiv0$ for $\mB \le \mB_c$ and obtain the straight interval for $0\le\mB<\mB_c$ in the bottom right plot of Fig. \ref{fig:vor}.}  of $\mB$. Note that for 
\bea
\mB = \mB_c + \d\mB\,, \quad 0<\d\mB\ll 1\,,
\eea
we have 
\bea
S_0 \approx 10.66 \ \d\mB\quad \text{and} \quad \int \limits_{\text{vor}}dC dL \approx 0.51(\d\mB)^2 \,.
\eea
Thus, $\eta_{\text{vor}}$ is of $\mO(\d\mB)$ near $\mB = \mB_c$. When $\mB$ is increasing, $\eta_{\text{vor}}$ is getting larger, which implies that the enlarged electromagnetic field triggers more vortical motions of charged particles. However, as $\mB$ increases enough, we have $\eta_{\text{vor}} \rightarrow 0.062$. In this case, the system is almost completely dominated by the Lorentz force, and the phase space gains an emergent symmetry, that is, $\mB \rightarrow \lm \mB, L \rightarrow L/\lm, C \rightarrow C/\lm^2$, where $\lm$ is the scaling factor.

The novel features related to a charged black hole discussed above might have some astrophysical significance. In the Gauss unit, for a particle of charge $q$ and mass $m$, there is
\bea
\mB = \f{1}{G} \f{q}{m} \f{Q}{M} \sim 3 \times 10^{22} \bigg(\f{q}{e}\bigg) \bigg( \f{m_p}{m} \bigg) \bigg(\f{Q}{\sqrt{G}M}\bigg) \, ,
\eea
where $e$ is the unit charge, $m_p$ is the ion's mass (the hydrogen nucleus), and $Q$ is the black hole charge. Astronomically, the black hole accretes hot plasma to the near-horizon region, forming a disk region near the equatorial plane. Moreover, abundant collisionless particles can be accreted or produced outside the disk region \cite{Ruffini:1975ne}, where they are accelerated by the magnetosphere and emit non-thermal synchrotron radiation. Note that the single-particle approximation only applies outside the disk region, thus only the vortical motions can produce non-thermal radiations in NHEK. Considering the radiations could escape to infinity, the signature of the charged particles moving vortically might be observed by the telescopes.

On the other hand, as the specific charge of electron is much larger than that of ions, so that $\mB_{\text{e}} \approx 2000 \mB_{\text{ion}}$. If $\mB_e \lesssim \mB_c$, there are only oscillatory motions and no non-thermal radiations. If $\mB_i \lesssim \mB_c < \mB_e$, only electrons can move vortically, while if $\mB_i > \mB_c$, both electrons and ions can move vortically. As an example, for a supermassive black hole with mass $M = 10^{10}M_{\odot}$, as long as the black hole charge $Q$ is larger than $1.4 \ \text{C}$, the radiations from near-horizon vortical electrons could be observed. However, the radiations from near-horizon vortical ions could be observable only when $Q > 2800 \ \text{C}$. 

Finally, we would like to comment on the extraction of rotational energy through electromagnetic fields. The Blandford-Znajek mechanism \cite{Blandford:1977ds} states that the energy of a Kerr black hole can be extracted through magnetic field torsion in the form of Poynting flux based on the force-free solution. On the contrary, unlike the force-free solution, the vacuum solution cannot generate an energy flux from the horizon, suggesting that energy extraction through the vacuum electromagnetic field is not feasible. Additionally, as the NHEK spacetime on which the present work based is not asymptotically flat, a more careful analysis is required to see if the vortical motion can be extended to the asymptotical flat region and check the efficiency of energy extraction.

\section{ Summary and discussion } \label{sec5}

In this work, we studied the motions of charged particles under MSEM fields in the NHEK geometry. Due to the enhanced symmetry, there are enough conserved quantities to simplify the equations of motion, which turn out be only a set of decoupled  first-order differential equations. Even though the radial motions are similar to the geodesics in NHEK,  the angular motions are changed significantly by the electromagnetic fields.  We investigated the motions of charged particles in two MSEM fields, the force-free field and the field in vacuum solutions. In the force-free case, there are stable equatorial motions and oscillatory motions, similar to the geodesics in NHEK. In the vacuum solution, we mainly focused on the case that $A_E=0$, which recovers the Wald potential in the NHEK geometry, and discussed  the motions of charged particles in detail.

The motions of charged particles under the Wald potential in NHEK geometry present some novel features. In the NHEK region, the Wald potential is dominated by an electromagnetic field  Eq.~\eqref{Wald}. In this case, the angular motions could be  classified into two types, one being the oscillatory motions which cross the equatorial planes continually, and the other being the vortical motions without crossing the equatorial plane.  We found the critical value of the electromagnetic parameter $\mB_c = \sqrt{3}$ above which the vortical motions would occur. Among the vortical motions, there is a special subclass of motions, in which the particles move  in a conical surface with fixed $\t$. Actually in the ``phase space" of the motions Fig.~\ref{fig:vor}, the vortical motions are surrounded by the conical motions and unstable equatorial motions. We calculated the conserved quantities of the conical motions and showed their changes with $\mB$ in the top right plot of the Fig.~\ref{fig:vor}.

In addition, since the electromagnetic field can be seen as a magnetosphere produced by a weakly charged black hole, we further discussed some astrophysical implications of $\mB_c$ and found that even a weak black hole charge might induce a significant difference between the behaviors of electrons and ions in NHEK, thus triggering relevant observational signatures.


We close this paper with some outlooks. On the one hand, as the first step to considering more realistic models, the discussion is limited to charged particle dynamics in the NHEK geometry with a weak electromagnetic field. It would be interesting to consider the effect of a strong field that affects the background geometry, like the near-horizon description of extreme Ernst–Wild solution and extreme MKN black holes \cite{Bicak:2015lxa}. On the other hand, it is of both theoretical and astrophysical importance to extend the study to the whole extreme Kerr throat, not just the NHEK region. It is also essential to investigate the observational signatures of charged particles inside or outside the Kerr throat.

\section*{Acknowledgments}
We are very thankful to Qiaomu Peng, Haopeng Yan for helpful discussions. The work is partly supported by NSFC Grant No. 12275004, 12205013 and 11873044. MG is also endorsed by ”the Fundamental Research Funds for the Central Universities” with Grant No. 2021NTST13.

\appendix

\section{Proof of $C+L^2 \geq 0$ in the Wald potential}\label{proof}
In this section we demonstrate that the constraint $\md = C+L^2 \geq 0$ holds for the motions in NHEK under the Wald potential Eq. (\ref{Wald}). The angular potential takes 
\be
\T(z) = \ma z^2 + \mb z +\mc
\ee
with 
\bea
\ma = 1-(L+\mB)^2/4\,, \quad\mb = -C - 3L^2/2 +\mB^2/2\,, \quad\mc = C + (L + \mB)(3L - \mB)/4-1\,.
\eea
On one hand, for the oscillating motions, we have
\be
\md > C+L^2-1 \geq C + \f{1}{4}(L + \mB)(3L - \mB)-1 = \mc \geq 0 \, ,
\ee
where we have used $(L+\mB)(3L-\mB) \leq 4L^2$. So we have $d > 0$ for the oscillating motions. Then, for the vortical motions, the constraint is 
\bea
\ma < 0\,,\quad\mc<0\,,\quad 0<\mb<-2\ma\,,\quad\mb^2-4\ma\mc >0\,.
\eea
Define $x_{\pm} = L \pm \mB$, thus
\bea\label{abcx}
\ma = 1-\f{1}{4}x_-^2  \, , \ \ \mb = \f{1}{2}x_+x_- - \md \, , \ \ \mc = \md - \f{1}{4}x_+^2-1 \, .
\eea
The expression of $x_{\pm}$ can be written as
\be\label{xpm}
x_+ = \pm \f{\mb + \md}{\sqrt{1-\ma}} \, , \ \ x_- = \mp 2\sqrt{1-\ma} \, .
\ee
Inserting Eq.(\ref{xpm}) into the third equation of Eq.(\ref{abcx}) to obtain
\bea\label{d}
\md^2 +\mS \md + \mT=0 \, ,
\eea
where 
\bea
\mS = 2(\mb+2\ma-2)\,,\quad\mT=\mb^2+4(1-\ma)(1+\mc)\,.
\eea
The roots of Eq.(\ref{d}) are
\be
\md_{\pm} = \f{1}{2}\big(-\mS \pm \sqrt{\mS^2-3\mT} \big) \,  .
\ee
Since $0<\mb<-2\ma$, one has $\mS <-4$. Moreover, the inequality $4\ma\mc<\mb^2<4\ma^2$ means $\ma<\mc<0$, thus we have
\bea
\mT = \mb^2-4\ma\mc+4(1-\ma+\mc) >4(1-\ma+\mc) >4>0\,.
\eea
Therefore, we have $|\mS| > \sqrt{\mS-4\mT}$ and the roots must satisfy
\be
\md_+ > \md_- >0 \, ,
\ee 
which means $d > 0$ for the vortical motions.

\end{document}